\documentclass[reprint, 12pt, 3p,times]{elsarticle}
\usepackage{amsmath}
\usepackage{amssymb,amsthm,graphicx}
\usepackage{tabularx}
\usepackage{multirow}
\usepackage{tabularx}
\usepackage{float}
\usepackage{stackengine}
\usepackage{BOONDOX-cal}
\usepackage{mciteplus}
\usepackage{bm}
\usepackage{natbib}

\usepackage[usenames, dvipsnames]{color}
\journal{Phys. Lett. A}
\date{\mute}
\usepackage{graphicx}
\graphicspath{{./figures/}}
\usepackage{epsfig}
\usepackage{float}
\usepackage[utf8]{inputenc}
\newfont{\logo}{logo10}

\newcommand{\bea}{\begin{eqnarray}}
\newcommand{\eea}{\end{eqnarray}}
\newcommand{\bes}{\begin{subequations}}
	\newcommand{\ees}{\end{subequations}}

\newcommand{\sech}{{\rm sech}}

\begin{document}

\begin{frontmatter}

\title{On the integrability aspects of nonparaxial nonlinear Schr\"odinger equation and the dynamics of solitary waves}

\author[KT]{K. Tamilselvan}\ead{tamsel.physics@gmail.com}
\address[KT]{Nonlinear waves Research Lab, PG \& Research Department of Physics, Bishop Heber college (Affiliated to Bharathidasan University), Tiruchirapppalli-620017, Tamil Nadu, India}
\author[KT]{T. Kanna\corref{Corresponding author}}
\cortext[Corresponding author]{Corresponding author}
\ead{kanna\_phy@bhc.edu.in}
\author[AG]{A. Govindarajan}\ead{govin.nld@gmail.com}
\address[AG]{Centre for Nonlinear Dynamics, School of Physics, Bharathidasan University, Tiruchirappalli - 620 024, Tamil Nadu, India}

\begin{abstract}
The integrability nature of a nonparaxial nonlinear Schr\"odinger (NNLS) equation, describing  the propagation of ultra-broad nonparaxial beams in a planar optical waveguide, is studied by employing the Painlev\'e singularity structure analysis. Our study shows that the NNLS equation  fails to satisfy the Painlev\'e test. Nevertheless, we construct one bright solitary wave solution for the NNLS  equation by using the Hirota's direct method. Also, we numerically demonstrate the stable propagation of the obtained bright solitary waves even in the presence of an external perturbation in a form of white noise. We then numerically investigate the coherent interaction dynamics of two and three bright solitary waves. Our study reveals interesting energy switching among the colliding solitary waves due to the nonparaxiality.
\end{abstract}


\end{frontmatter}

\section{Introduction}
The advent of the universal nonlinear Schr\"odinger (NLS) equation  in nonlinear optics has opened an  avenue to explore various nonlinear waves like solitons, breathers, rogue waves, shock waves, vortices and so on \cite{Kivshar2003, Ablowitz,Kharif}.  This ubiquitous model can be derived  from the famous Maxwell's equations  by employing the so-called slowly varying envelope approximation (SVEA) alias paraxial approximation (PA), which is justified only if the scale of spatial and temporal variations is larger than the optical wavelength and optical cycle, respectively. Under this approximation, the second-order derivative of the normalized envelope field, with respect to its longitudinal (propagation) co-ordinate, can be ignored. From a practical point of view, the SVEA holds good when the optical modes are propagating along (or at near-negligible angles with) the reference axis with its pulse/beam width being greater than the carrier wavelength. Though the NLS equation naturally describes the pulse propagations in optical fibers \cite{Gadi} with such limitations, the pulses encounter a catastrophic collapse when higher order traverse dimensions are included \cite{Feit,Kelly}. It should be noted that the inclusion of nonparaxiality (or spatial group velocity dispersion (S-GVD)) has led to the stable propagation of localized pulses  even in  higher dimensional NLS equations \cite{Feit}.

In addition to nonlinear optics, this S-GVD or nonparaxial effect appears naturally in the  dynamics of exciton-polaritons in a waveguide of semiconductor material such as ZnCdSe/ZnSe superlattice \cite{Biancalana}. The underlying governing equation is the NLS equation with spatial dispersion term. This system is formally equivalent to the nonparaxial NLS equation (also referred as nonlinear Helmholtz  (NLH) equation) which is routinely used to study nonparaxial localized modes in optical wave guides \cite{Christ1}.  In the earlier work of Lax \emph{et al.,}  \cite{Louis}, it was attempted to investigate the nonparaxial effect by means of expanding field components as a power series in terms of a ratio of the beam diameter to the diffraction length. Following this  work, many studies have been carried out to investigate the dynamics of nonparaxiality in various optical settings like  nonparaxial accelerating beams  \cite{Zhang}, optical and plasmonic sub-wavelength nanostructures devices \cite{Gramotnev,Liu,Gorbach}, and in the design of Fresnel type diffractive optical elements \cite{Nguyen}.

Furthermore, the propagation of nonparaxial solitons has  stimulated extensive studies in different nonlinear optical settings such as Kerr media \cite{Christ1}, cubic-quintic media \cite{Bistable}, power-law media \cite{powerlaw}, and saturable nonlinear media \cite{Saturable}.  The soliton theory has also been formulated in the NLH equation with distinct nonlinearities based on relativistic and pseudo-relativistic framework \cite{Christian,christian0,christian1}. The coupled version of the NLH equation has been studied to explore various kinds of nonlinear waves, including elliptic waves and solitary waves \cite{Blair,tamil}. Recently, the study of nonparaxiality has been extended to the  intriguing area of $\mathcal{PT}$-symmetric optics \cite{Huang}. Also, quite recently, the present authors have done a systematic analysis of the  modulational instability  for the cubic-quintic NLH equation and reported various interesting chirped elliptical and solitary waves with nontrivial chirping behavior \cite{Tamil1}.

In nonlinear dynamical systems, the challenging problem is to identify  new nonlinear integrable/nearly integrable models. This has an important consequence for exploring nontrivial localized nonlinear waves with intriguing dynamical features in different physical media \cite{Ablowitz, Boris1,Boris2}. Moreover,  investigations of the integrability nature of dynamical systems have been extended to multiple areas of physics, including fluid dynamics, nonlinear optics, Bose-Einstein condensates, bio-physics and so on.  Specifically, one can verify the integrability nature of a nonlinear dynamical equation by using a powerful mathematical tool, namely, Painlev\'e analysis \cite{Tabor1, Tabor}.
The Painlev\'e analysis is a potential tool among many integrability indicators such as the linear eigenvalue problem, bilinear transformation, B\"acklund transformation, Lax-pair method, and inverse scattering method \cite{Ablowitz, Lakshmanan}. Through the Painlev\'e analysis, the integrability nature has been tested for various nonlinear models \cite{Ablowitz1980, LAKSHMANAN19931, Ramani1989}. As mentioned earlier, the NNLS equation  can serve as a fertile platform for studying dynamics of a wide range of physical systems. In a recent work, the symmetry reductions of the NNLS equation have been obtained by the Lie symmetry analysis \cite{Sakkaravarthi2018}.
 However,  the integrability nature of this NNLS equation is yet to be investigated. One of the objectives of  this paper is to inspect the integrability nature of the following dimensionless NNLS equation.
\begin{equation}\label{NLH}
i\frac{\partial \mathbf{\mathbf{\Psi}}}{\partial z}+\kappa~ \frac{\partial^2 \mathbf{\mathbf{\Psi}}}{\partial z^2}+\frac{1}{2}\frac{\partial^2 \mathbf{\mathbf{\Psi}}}{\partial t^2}+\gamma|\mathbf{\mathbf{\Psi}}|^2 \mathbf{\mathbf{\Psi}}=0,
\end{equation}
where  $\mathbf{\Psi}$ is the normalized complex envelope field and  normalized space $z$ and retarded time $t$ are expressed as $Z/L_{D}$ and $T/T_{0}$,  in which the dispersion length $L_{D}$ is determined by $T_{0}^{2}/|\beta_{2}|$.  The parameters $\beta_{2}$ and $T_{o}$ account for group velocity dispersion (GVD) and input pulse width, respectively. The term $\kappa$ refers to nonparaxial parameter which ranges from $10^{-2}$ to $10^{-4}$ with $\kappa=1/ 2 k_{0} L_{D}$  (where $k_{0}=2\pi n_{0}/\lambda$ stands for wavenumber, in which $n_{0}$ is refractive index) \cite{Christian}. Also, the term $\gamma$ indicates the Kerr nonlinearity co-efficient. In the limit, $\kappa \rightarrow 0$ the equation (\ref{NLH}) reduces to the standard  NLS equation.

The paper is organized as follows. In Sec. 2, we carry out the investigation of the integrability nature of the NNLS equation with the aid of Painlev\'e analysis which consists of three steps, namely calculating the leading-order equation,  finding the resonance values and verifying the existence of sufficient number of arbitrary functions without the movable critical singularity manifolds. In Sec. 3, we obtain the bright solitary wave solution by employing the Hirota's bilinearization method. Following that, the stability of the bright solitary wave solution in the presence of  external perturbation is examined by numerical simulation in Sec. 4. In addition, the interaction of nonparaxial solitary waves is analyzed by executing split step Fourier (SSF) method. Finally, we conclude our findings in Sec. 5.

\section{Painlev\'e Singularity Structure Analysis}
In order to apply the Painlev\'e singularity structure analysis to equation (\ref{NLH}), we consider the dependent variable and its complex conjugate as $\mathbf{\mathbf{\Psi}}=r,\,\,\mathbf{\mathbf{\Psi}}^{*}=s.$ Then the equation \eqref{NLH} and its complex conjugate equation can be rewritten as,
\bes\label{3}\bea
ir_{z}+\kappa r_{zz}+\frac{1}{2}r_{t t}+\gamma (r^2 s)=0\,,\label{3a}\\
-is_{z}+\kappa s_{zz}+\frac{1}{2}s_{t t}+\gamma (s^2 r)=0.\,\label{3b}
\eea\ees
The singularity structure analysis of the above equations \eqref{3a}-\eqref{3b} is carried out by seeking the following generalized Laurent series expansion for the dependent variables in the neighbourhood of the non-characteristic singular manifold $\phi(z,t)=0$ with non-vanishing derivatives i.e., $\phi_{z}(z,t)\ne0$ and  $\phi_{t}(z,t)\ne0$:
\bes\label{4}\bea
r=\phi^{\alpha}\displaystyle\sum_{j=0}^{\infty}r_{j}(z,t)~\phi^{j},\,\,r_{0}\ne0,\label{4a}\\
s=\phi^{\beta}\displaystyle\sum_{j=0}^{\infty}s_{j}(z,t)~\phi^{j},\,\,s_{0}\ne0,\label{4b}
\eea \ees
where $\alpha$ and $\beta$ are  integers yet to be determined.
Next, in order to analyze the leading order solution, we restrict the above series as, $r=r_{0}~\phi^{\alpha}$ and $s=s_{0}~\phi^{\beta}$. By using these relations in equation~(\ref{3}) and balancing the most dominant terms, the unknown values $\alpha$ and $\beta$ are determined as, $\alpha=-1,$ and $~\beta=-1,$ accompanied by the following condition
\bea\label{6}
2\kappa\phi_{z}^{2}+\phi_{t}^{2}=-\gamma(r_{0} s_{0}).
\eea
In equation \eqref{6}, out of two functions $r_{0}$ and $s_{0}$ , one is arbitrary.

Next, the resonances (powers at which arbitrary functions can enter into the Laurent series (\ref{4})) are obtained by  determing the values of $j$  upon substitution of the following equations
\bes\label{7}\bea
r=r_{0}~\phi^{-1}+\cdots+r_{j}~\phi^{j-1},\,\,\\
s=s_{0}~\phi^{-1}+\cdots+s_{j}~\phi^{j-1}\,\,\,\,
\eea \ees
 into equations \eqref{3}. By collecting the coefficients of  $\phi^{j-3}$, we get
\bea\label{8}
\left(\begin{array}{cc}(j^{2}-3j)\delta+\gamma(r_{0}s_{0}) &\gamma r_{0}^2  \\
	\gamma s_{0}^2 & (j^{2}-3j)\delta+\gamma(r_{0}s_{0})
\end{array}
\right)\left( \begin{array}{c} r_{j}\\s_{j}\\ \end{array} \right)=0,\nonumber\\
\eea
where $\delta=\kappa\phi_{z}^{2}+\frac{1}{2}\phi_{t}^{2}$. By setting the above matrix determinant to be zero, we obtain a quartic equation for $j$ as follows
\bea\label{9}
j^4-6 j^{3}+5 j^{2}+12j=0.
\eea
The roots of equation (\ref{9}) are the resonance values and are found to be $j=-1,0,3,4$, where the resonance value $j = -1$ denotes the arbitrariness of the singular manifold $\phi(z, t)$. Except this, all other resonance values are positive as required by the Painlev\'e test.
\subsection{Arbitrary Analysis}\label{arbitray_anlysis}
The third step is to examine the existence of sufficient number of arbitrary functions at these resonance values without introducing movable critical singular manifolds of the singularity structure analysis. To this end, we expand the dependent variables as follows:
\bes\label{Eq10}\bea
r=\frac{r_{0}}{\phi}+r_{1}+r_{2}\phi+r_{3}\phi^{2}+r_{4}\phi^{3},\label{Eq10a}\\
s=\frac{s_{0}}{\phi}+s_{1}+s_{2}\phi+s_{3}\phi^{2}+s_{4}\phi^{3}.\label{Eq10b}
\eea\ees
Then, substituting the above equations \eqref{Eq10} in equations \eqref{3} and collecting the co-efficients at various orders of $\phi$, one can study the arbitrariness of the singularity.

First, collecting the terms at the order $\phi^{-3}$ which corresponds to the resonance value $j=0$, we obtain
\bea
2 \kappa \phi_{z}^2+\phi_{t}^{2}=-\gamma(r_{0} s_{0}).
\eea
This equation is exactly the same as the leading order equation (\ref{6}).

Second, collecting the coefficients at the order $\phi^{-2}$, we obtain the following equations which are expressed in  matrix form as

	\bea\label{10}
	\left(
	\begin{array}{cc}
		-2(1+2\kappa \rho_{z}^{2})& \gamma r_{0}^{2}\\
		\gamma s_{0}^{2} & -2(1+2\kappa \rho_{z}^{2}) \\
	\end{array}
	\right)\left( \begin{array}{c} r_{1}\\s_{1}\\ \end{array} \right)=-\left(
	\begin{array}{c}
		i r_{0}\rho_{z}+\kappa r_{0}\rho_{zz}+2\kappa r_{0,z}\rho_{z}  \\
		-i s_{0}\rho_{z}+\kappa s_{0}\rho_{zz}+2\kappa s_{0,z}\rho_{z}\\
	\end{array}
	\right).
	\eea

Here, we have used the Kruskal ansatz of the form $\phi(z,t)=t+\rho(z)$, with $\rho(z)$ being an arbitrary analytic function to simplify the calculations and the $r_{j}$ and $s_{j}$ are functions of $z$ only. We obtain the following expressions for $r_{1}$ and $s_{1}$ from the above equation (\ref{10}).

\bes\label{10a}\bea
	r_{1}=\frac{1}{3(1+2\kappa \rho_{z}^{2})^{2}}[(i r_{0}\rho_{z}+\kappa r_{0}\rho_{zz}+2\kappa r_{0,z}\rho_{z})( 2(1+2\kappa \rho_{z}^{2}))+ \gamma r_{0}^{2}(-i s_{0}\rho_{z}+\kappa s_{0}\rho_{zz}+2\kappa s_{0,z}\rho_{z})],\,\,\,\,\\
	s_{1}=\frac{1}{3(1+2\kappa \rho_{z}^{2})^{2}}[(2(1+2\kappa \rho_{z}^{2}))(	-i s_{0}\rho_{z}+\kappa s_{0}\rho_{zz}+2\kappa s_{0,z}\rho_{z})+\gamma s_{0}^{2} (i r_{0}\rho_{z}+\kappa r_{0}\rho_{zz}+2\kappa r_{0,z}\rho_{z})].\,\,\,\,\,
\eea	\ees

Thus, there is no arbitrary function at this order.

Similarly, collecting the coefficients at the order $\phi^{-1}$, we obtain

	\bea\label{11}
	\left(
	\begin{array}{cc}
		-2(1+2\kappa \rho_{z}^{2})& \gamma r_{0}^{2}\\
		\gamma s_{0}^{2} & -2(1+2\kappa \rho_{z}^{2}) \\
	\end{array}
	\right)\left( \begin{array}{c} r_{2}\\s_{2}\\ \end{array} \right)=-\left(
	\begin{array}{c}
		i r_{0,z}+\kappa r_{0,zz}+\gamma(r_{1}^{2}s_{0}+2 r_{0} r_{1}s_{1})\\
		-i s_{0,z}+\kappa s_{0,zz}+\gamma(s_{1}^{2}r_{0}+2 s_{0} r_{1}s_{1})\\
	\end{array}
	\right).
	\eea

By solving the above set of algebraic equations, we find  that $r_{2}$ and $s_{2}$ can be expressed in terms of $r_{0}$ and $s_{0}$ as below

	\bes\label{13}\bea
	r_{2}=\frac{1}{3(1+2\kappa \rho_{z}^{2})^{2}}[2(i r_{0,z}+\kappa r_{0,zz}+\gamma (r_{1}^{2}s_{0}+2 r_{0} r_{1} s_{1}))(1+2 \kappa\rho_{z}^{2})\nonumber\\
	+\gamma r_{0}^{2}(-i s_{0,z}+\kappa s_{0,zz}+\gamma (s_{1}^{2}r_{0}+ 2s_{0} r_{1} s_{1}))],\\
	s_{2}=\frac{1}{3(1+2\kappa \rho_{z}^{2})^{2}}[2(-i s_{0,z}+\kappa s_{0,zz}+\gamma (s_{1}^{2}r_{0}+ 2 s_{0} r_{1} s_{1}))(1+2 \kappa\rho_{z}^{2})\nonumber\\
	+\gamma s_{0}^{2}(i r_{0,z}+\kappa r_{0,zz}+\gamma (r_{1}^{2}s_{0}+2 r_{0} r_{1} s_{1}))],
\eea	\ees

\noindent where the expressions for $r_{1}$ and $s_{1}$ are as given in equations \eqref{10a}. The above equations \eqref{13} indicate that $r_{2}$ and $s_{2}$ are not arbitrary functions. Further, collecting the coefficients at the order $\phi^{0}$ corresponding to the resonance value $j=3$, we obtain,
	\bes\label{14}\bea
	s_{0} r_{3}+r_{0}s_{3}=-\frac{1}{\gamma r_{0}}[ir_{1,z}+\kappa r_{1,zz}+i r_{2}\rho_{z}+\kappa r_{2} \rho_{zz}\label{14a}\nonumber\\
	+2 \kappa r_{2,z}\rho_{z}+\gamma r_{1}(r_{1} s_{1}+2 r_{0} s_{2}+2r_{2} s_{0})+2 \gamma r_{0} r_{2}s_{1}],\\
	s_{0} r_{3}+r_{0}s_{3}=-\frac{1}{\gamma s_{0}}[-is_{1,z}+\kappa s_{1,zz}-i s_{2}\rho_{z}+\kappa s_{2} \rho_{zz}\label{14b}\nonumber\\
	+2 \kappa s_{2,z}\rho_{z}+\gamma s_{1}(r_{1} s_{1}+2 r_{0} s_{2}+2r_{2} s_{0})+2 \gamma s_{0} s_{2}r_{1}].
\eea	\ees

By carefully analyzing the right hand sides of expressions (\ref{14}) by symbolic computation, we note that they become non-identical except for the choice $\kappa = 0,$ which corresponds to the result of the standard integrable NLS equation. This clearly indicates the violation of arbitrariness for the resonance $j=3$, as there is no any arbitrary function. Hence  the NNLS equation \eqref{NLH} fails to satisfy the Painlev\'e property at this stage.

Finally, we move on to collect the coefficients at the order $\phi^{1}$ and one obtains
\bea\label{15}
\left(
\begin{array}{cc}
	(1+2\kappa \rho_{z}^{2})& \gamma r_{0}^{2}\\
	(1+2\kappa \rho_{z}^{2})& \gamma r_{0}^{2}\\
\end{array}
\right)\left( \begin{array}{c} r_{4}\\s_{4}\\ \end{array} \right)=A,
\eea

where the column matrix  $A$ is given by

	\bea\label{16}
	A=-\left(
	\begin{array}{c}
		i (r_{2,z} +2 r_{3} \rho_{z}) + \kappa( r_{2,zz} +2 r_{3} \rho_{zz}+4r_{3,z} \rho_{z})+\gamma(r_{1}^{2}s_{2}+2 r_{0} r_{2} s_{2} + 2 r_{1} s_{1} r_{2}\nonumber\\+2 r_{1} r_{3} s_{0}+ 2 r_{0} r_{3} s_{1}+ 2 r_{0} r_{1} s_{3}+ r_{2}^{2} s_{0}) \nonumber\\
		-\frac{r_{0}^{2}}{(1+2\kappa \rho_{z}^{2})}( i (s_{2,z} +2 s_{3} \rho_{z}) + \kappa( s_{2,zz} +2 s_{3} \rho_{zz}+4s_{3,z} \rho_{z})+\gamma(s_{1}^{2}r_{2}+2 s_{0} s_{2} r_{2} + 2 s_{1} r_{1} s_{2}\nonumber\\+2 s_{1} s_{3} r_{0}+ 2 s_{0} s_{3} r_{1}+ 2 s_{0} s_{1} r_{3}+ s_{2}^{2} r_{0}) )
	\end{array}
	\right)
	\eea

As before, here also a rigorous analytical calculation involving symbolic computation shows that the above two equations remain distinct as long as $\kappa$ is non-zero. However, they become identical for $\kappa =0$, as expected. Thus, due to the failure of existence of sufficient number of arbitrary functions (see equations  \eqref{14} to \eqref{16}), we conclude that the NNLS equation (\ref{NLH}) is not free from movable critical singular manifolds. The above singularity structure analysis clearly indicates that  the NNLS equation  (\ref{NLH}) is not integrable in the Painlev\'e sense.
\section{ Solitary wave solutions for the NNLS equation}
\subsection{Hirota's Bilinearization method}
As established in the previous section, the NNLS equation \eqref{NLH} fails to satisfy the Painlev\'e test for integrability. Hence, one has to consider quasi-analytical methods or numerical analysis to find special solutions in the equation \eqref{NLH} \cite{Bountis1,Bountis2,Bountis3}. However, we here attempt to find special solutions in  equation \eqref{NLH} by using the well-known Hirota's bilinearization method, in spite of the equation \eqref{NLH} being non-integrable. The NNLS equation ~\eqref{NLH} is expressed in a bilinear form by employing the following transformation
\bea \label{bilinearform}
\mathbf{\mathbf{\Psi}}=\frac{g(z,t)}{f(z,t)},
\eea
where $g$ and $f$ are complex and real functions, respectively, and  $*$ indicates complex conjugation. The resulting bilinear equations are
\bes\label{bilinear}\bea
(i D_{z}+\kappa D_{z}^2+ \frac{1}{2} D_{t}^{2})(g\cdot f)=0,\\
(\kappa D_{z}^2+ \frac{1}{2} D_{t}^{2})(f\cdot f)=\gamma(g g^{*}).
\eea\ees
One can obtain the single solitary wave solution by expression
$g=\chi g_{1}$, and $f=1+\chi^2 f_2$ in equation (\ref{bilinear}), where $\chi$ is a formal expression parameter. Solving the resulting linear partial differential equations (\ref{bilinear}) at various orders of $\chi$ recursively, we obtain 
\bea\label{21}
\mathbf{\Psi}=\Delta~e^{i \eta_{1i}}~\sech \left(\eta_{1r}+\frac{R}{2}\right).
\eea
Here
\bes\bea
&&\eta_{1r}=a_{r}t+b_{r}z,\, \eta_{1i}=a_{i}t+b_{i}z+\theta,\,a_{i}=\small{\sqrt{-b_{i}+\kappa (b_{r}^{2}-b_{i}^{2})\pm\sqrt{(b_{r}^{2}+b_{i}^{2})[1+2 \kappa b_{i}+\kappa^{2}(b_{r}^{2}+b_{i}^{2})]}}},\nonumber\\
&&a_{r}=-\frac{b_{r}(2 \kappa b_{i}+1)}{a_{i}},\,\theta=\tan^{-1}\left(\frac{\alpha_i}{\alpha_r}\right),\,\,R=2\log{\sqrt{\frac{\gamma (\alpha \alpha^{*})}{(8\kappa b_{r}^{2}+4 a_{r}^{2})}}},\,\,\,
\eea
where $a_r$, $a_i$, $b_r$, $b_i$, $\alpha_r$, and $\alpha_i$ are real parameters. By direct substitution, we have also verified that the solution (\ref{21}) indeed satisfies the NNLS equation (\ref{NLH}).
This one bright solitary wave (\ref{21}) is characterized by four arbitrary real parameters $b_{r}$, $b_{i}$, $\alpha_r$ and $\alpha_i$. The amplitude  and velocity of one bright solitary wave (\ref{21}) can be expressed as
\bea
&&\Delta=\frac{\alpha}{2~e^{\frac{R}{2}}}=\frac{4 b_{r}\sqrt{\left[2\kappa a_{i}^{2}+(1+2\kappa b_{i}^{2})^{2}\right]} }{\sqrt{\gamma} a_{i}},\,\, \text{and}\,\, v=\frac{a_{i}}{(-2 \kappa b_{i}-1)},
\eea\ees
 respectively. Also, the phase part of the solitary wave is given by $a_{i}(t+\frac{b_{i}}{a_{i}}z)$. Here, the amplitude,  velocity and phase of the bright solitary wave are affected significantly by the nonparaxial effect due to the explicit appearance of the nonparaxial parameter $\kappa$ in their corresponding expressions. Note that, the  solution (\ref{21}) reduces to the standard  NLS soliton in the paraxial limit (i.e., when $\kappa  \rightarrow 0$). So, one can conclude that the nonparaxial parameter influences all the physical parameters of bright solitary wave of equation (\ref{NLH}). This is one of the distinct features of the  obtained solitary wave solution (\ref{21}).  tried to extend the above bilinearization procedure to obtain general two-soliton solution, but unsuccessful. This suggests that the three soliton solution of NNLS system \eqref{NLH} with a sufficient number of parameters does not exist. This conclusion is in support of the Painlev\'e analysis carried out in the previous section \eqref{arbitray_anlysis} showing the NLS system to be non-integrable.
 \begin{figure}[htb]
 	\centering\includegraphics[width=0.5\linewidth]{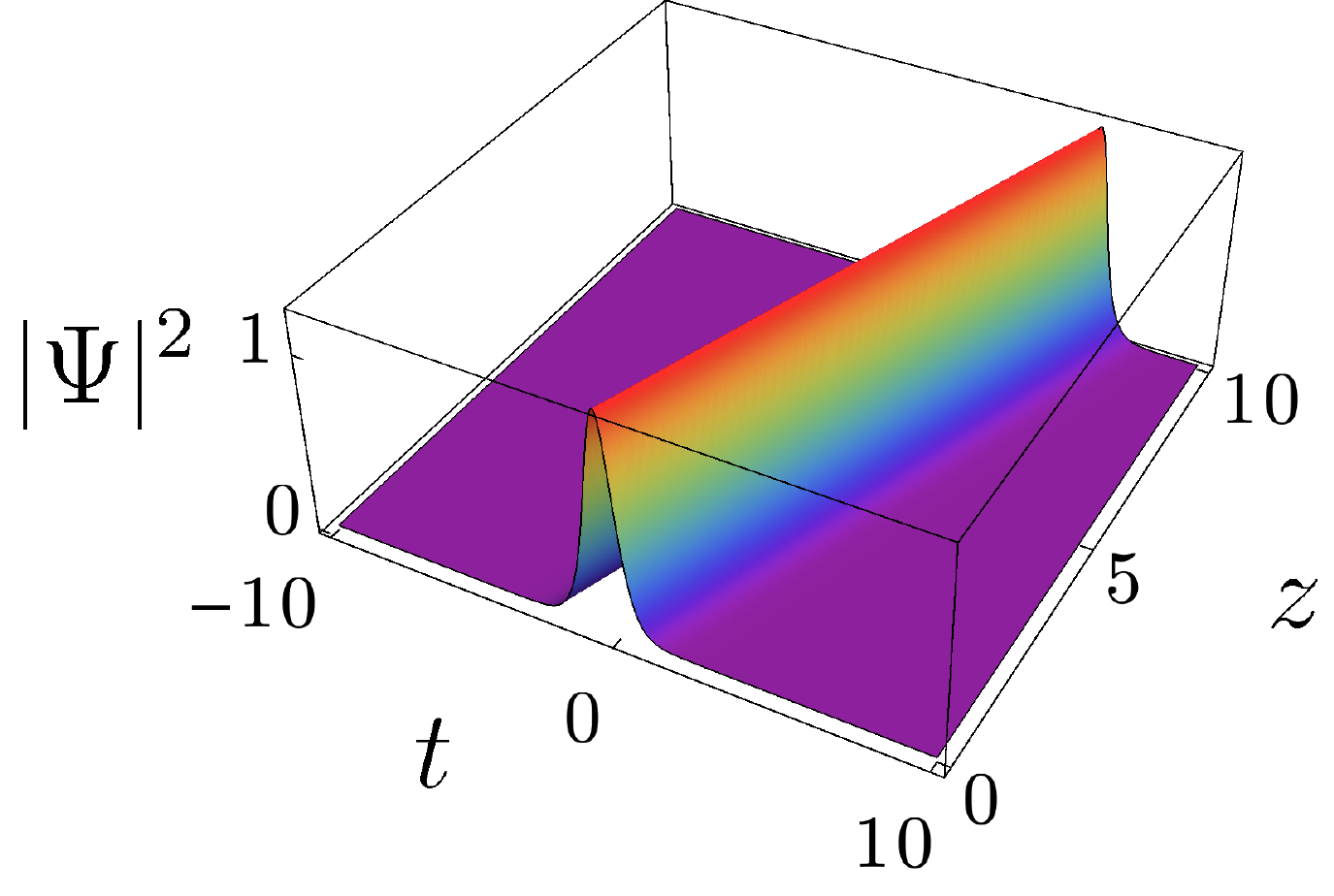}
 	\caption{Propagation of the bright one solitary wave for the parametric choice $b_{r}=b_{i}=\alpha_{1}=1$, $\kappa=0.01$, and $\gamma=2$.} \label{Figure02}
 \end{figure}
\begin{figure}[htb]
	\centering\includegraphics[width=0.4\linewidth]{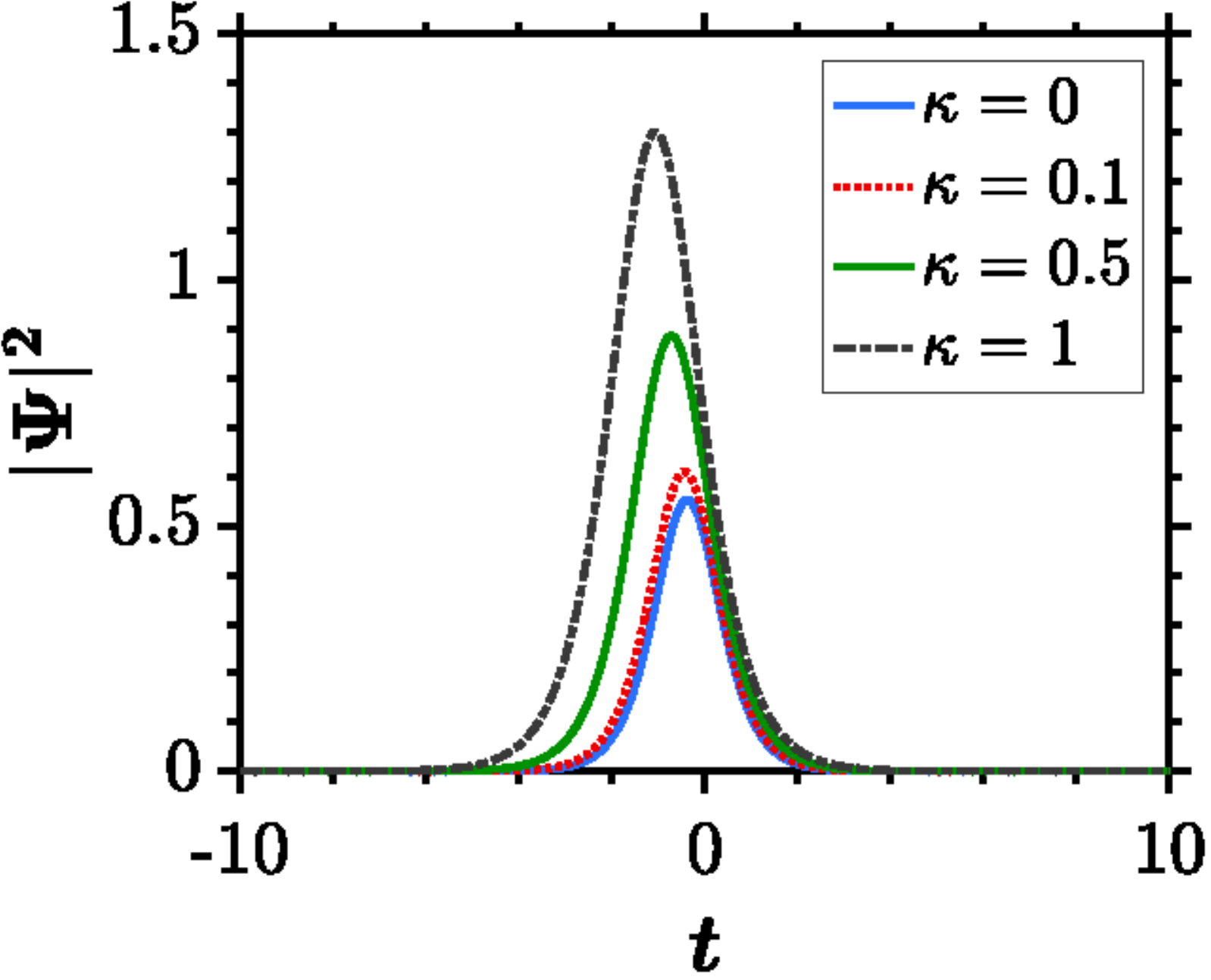}
		\caption{ Plot shows the bright one solitary waves for different values of $\kappa$ parameter. The parameters are assigned as $b_{r}=\alpha_{1}=1$, $b_{i}=0.1$, $\gamma=2$, and $z=0$.} \label{Figure2}
\end{figure}
\begin{figure}[t]
	\centering\includegraphics[width=0.4\linewidth]{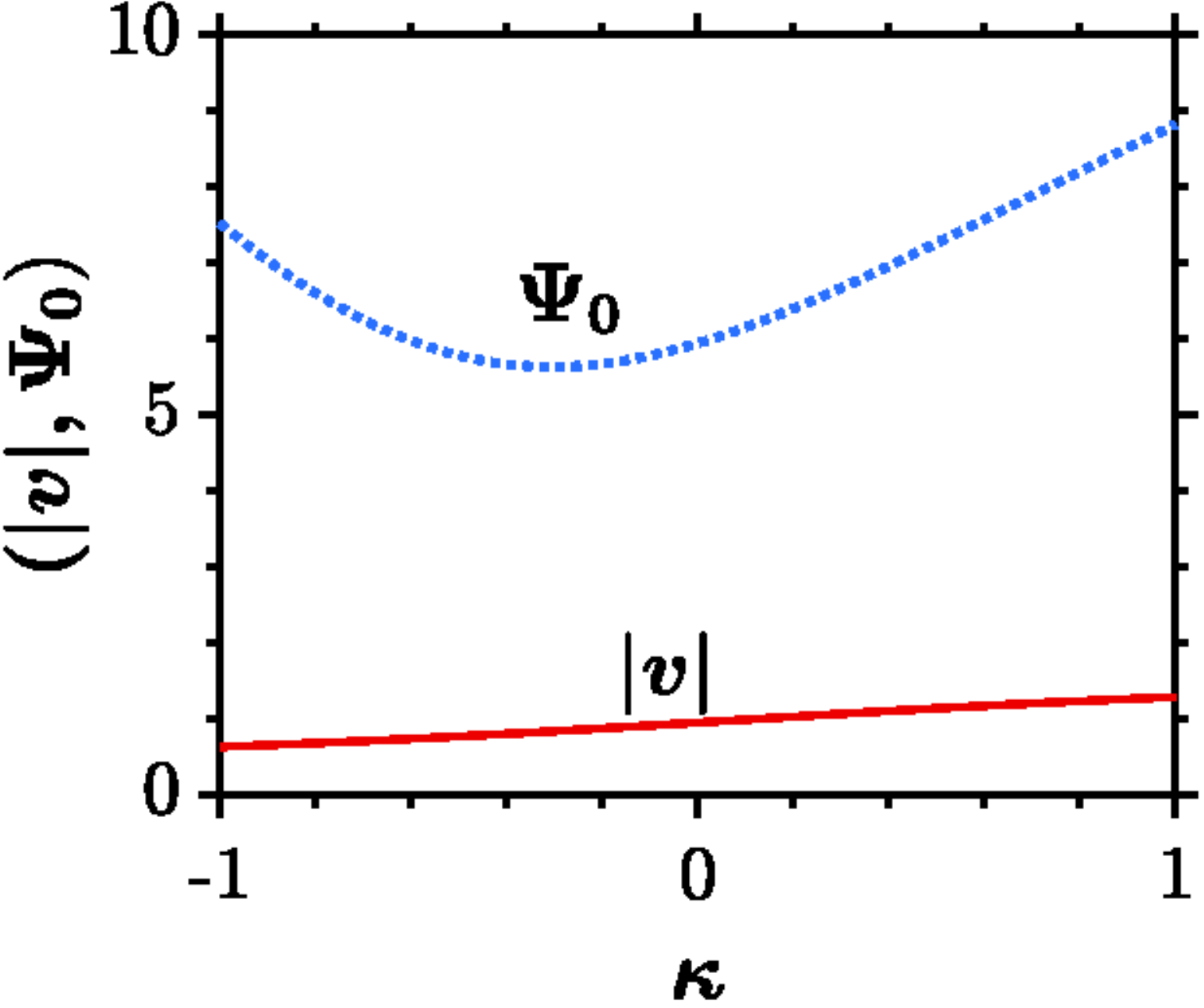}
	\caption{ Plot depicts the speed  and amplitude of the bright solitary wave as a function of the nonparaxial parameter $\kappa$. The parameters are same as given in Fig.~\ref{Figure2} } \label{Figure3}
\end{figure}
\begin{figure}[t]
	\centering
	\bottominset{(a)}{\centering\includegraphics[width=0.4\linewidth]{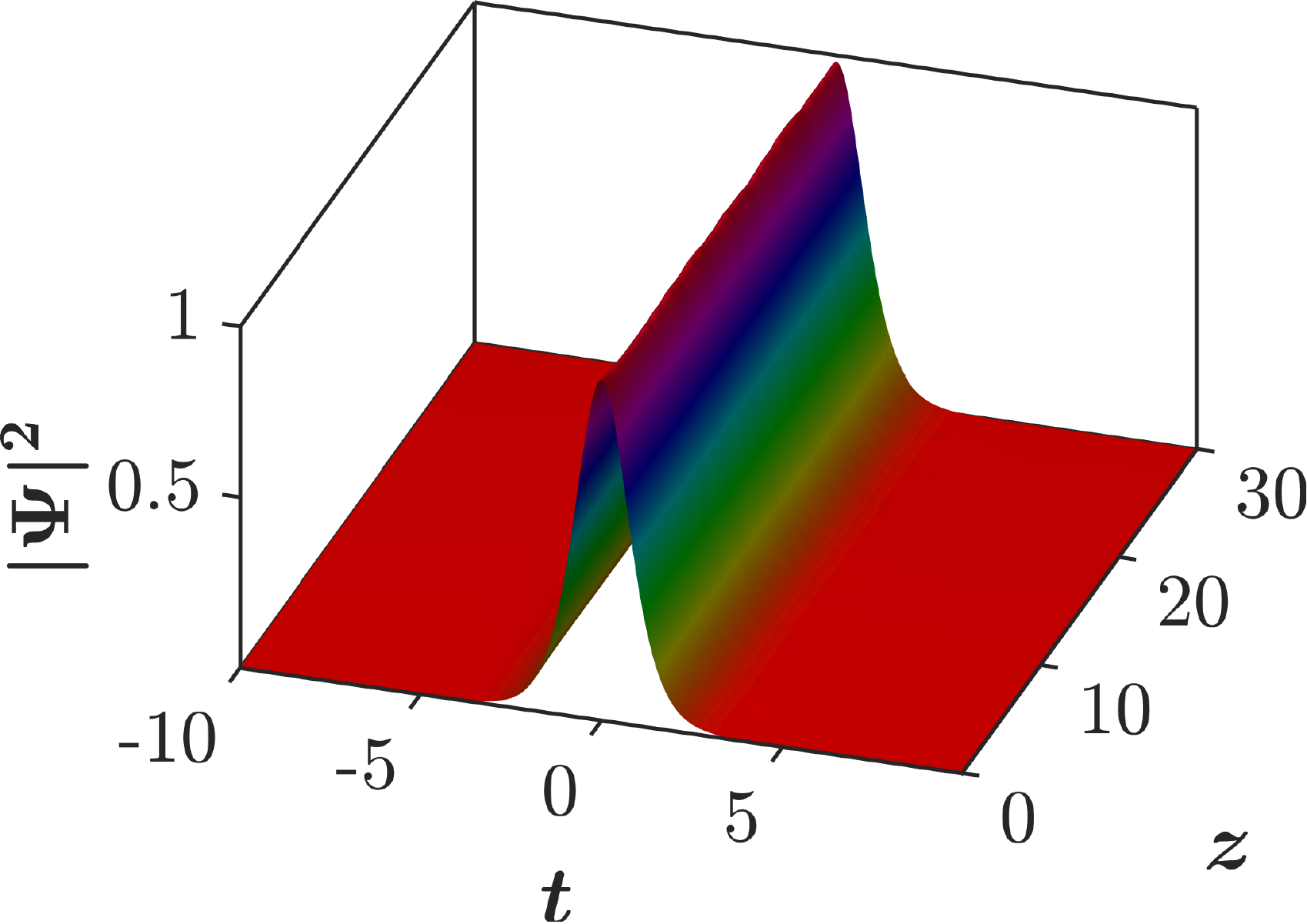}}{-0.2in}{-0.1in}
	\bottominset{(b)}{\centering\includegraphics[width=0.4\linewidth]{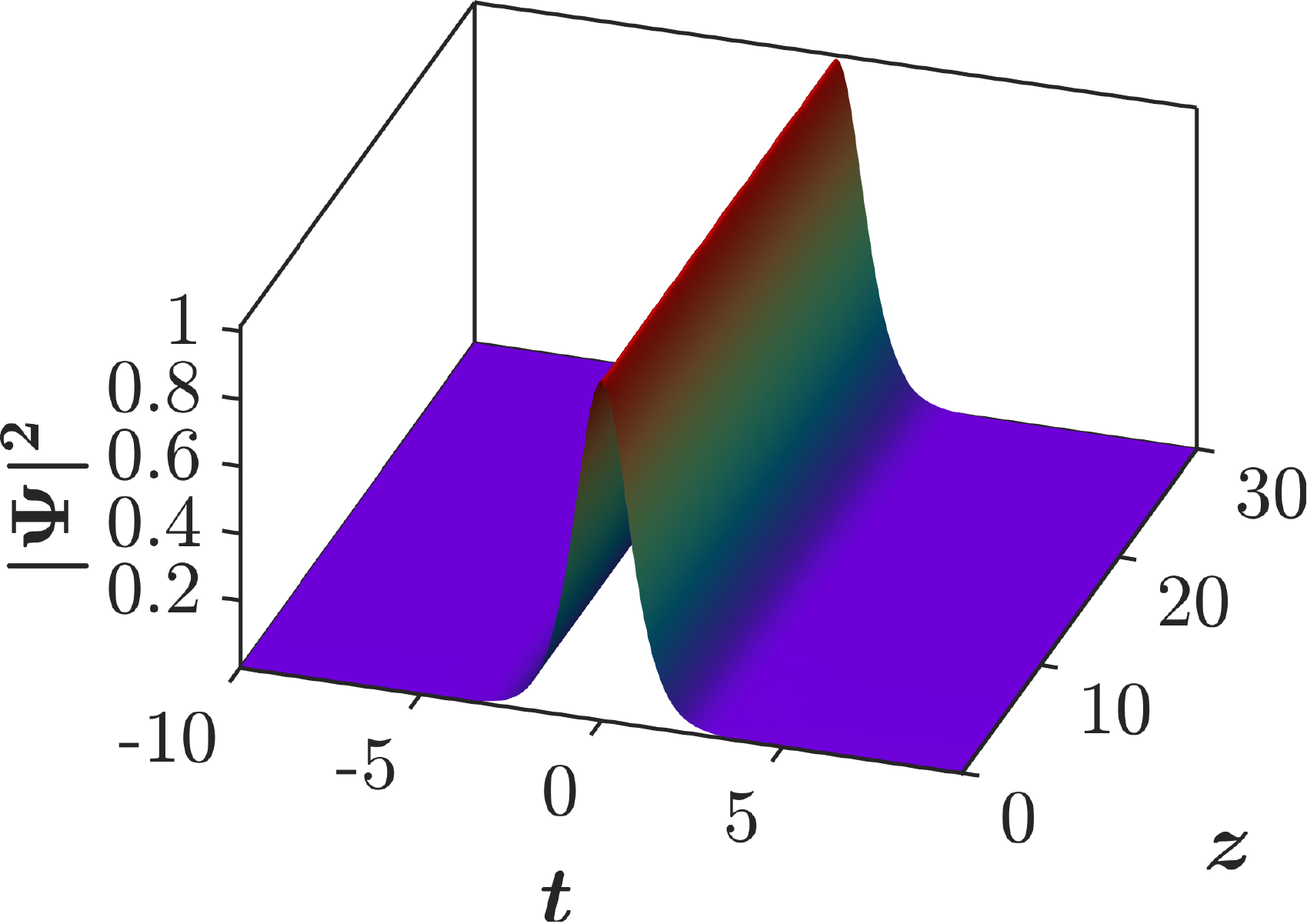}}{-0.2in}{-0.1in}
	\caption{Numerical evolution of stable bright solitary waves, in the absence of perturbation (a) and in the presence of white noise $10\%$ (b). The parameters are the same as in Fig.~\ref{Figure02}}\label{Numerical5}
\end{figure}

First, we show the propagation of the one bright solitary wave as in Fig.~\ref{Figure02} which mimics the typical soliton propagation in integrable systems. Then, in order to reveal the impact of nonparaxiality on the bright solitary wave,  we display the intensity plots of the bright solitary wave for different values of the nonparaxial parameter $\kappa$ in Fig.~\ref{Figure2}. In the  absence of the nonparaxial parameter (i.e. $\kappa=0$), it retraces the standard intensity profile as that of the NLS equation (solid black curve). Upon the onset of the nonparaxial parameter $\kappa$,  the bright solitary wave undergoes significant changes, not only in its amplitude and width but also in its central position. These are signatures of the nonparaxiality \cite{christian0,christian1}. The influence of  the nonparaxial parameter on physical quantities such as amplitude and speed of the bright solitary wave is presented  in Fig.~\ref{Figure3}. It clearly shows that the increase in the nonparaxial parameter enhances the speed of the bright solitary wave. For the $\kappa$ values lying in the window [-1,1], the amplitude decreases until $\kappa$ becomes zero and then it starts to increase.

We have also investigated the stable dynamics of obtained bright solitary wave of the NNLS equation by employing the split-step Fourier method based on Feit-Flock algorithm \cite{Feit}. To do so, we add a random uniform white noise as a perturbation at a rate of 10$\%$   in the initial solution of bright solitary wave solution \eqref{21}  \cite{Govind}.  Figure~\ref{Numerical5} demonstrates that the solitary  pulse remains stable for a long propagation distance which is quite larger for optical waveguides, without (see Fig. \ref{Numerical5}(a)) and with noise (see Fig. \ref{Numerical5}(b)). Hence the numerical evolution clearly demonstrates that the pulse is robust against small perturbations in the form of uniform white noise for the given system parameters.
\section{Scattering dynamics of bright solitary waves in the NNLS system}
Interaction of solitary waves is a key feature that determines their potential applications in nonlinear optical systems. It is interesting to study the interaction between two solitons by launching the soliton pulses far enough from each other, at least with a separation distance around ten times of their pulse-width \cite{Blow,Desem}. The implication of such criteria has really helped to overcome multiple issues like pulse distortion, deteriorations of the data transmission and synchronization in the high-bit-rate systems. In general, interaction of solitons can be classified into two main categories as coherent and incoherent  based on their relative phases \cite{TSANG,Lai}. In practice, the coherent type of interactions takes place when the interference effects between the overlapping beams are taken into account. It requires  the medium to respond instantaneously. On the contrary, incoherent interactions  exist when the time response varies slower than the relative phase between solitons. Ultimately, solitons experience periodic collapse with neighboring solitons. It must be hence noted that the incoherent interactions are undesirable in the practical viewpoint \cite{Stegeman1518}.

Interaction of various types of solitons has been intensively discussed both  from experimental and theoretical aspects \cite{Kivshar2003}. In particular, these studies considered  interaction between solitons/solitary waves in the NLS and NLS-like equations \cite{Boris3,Shalaby,TrikiT12,Triki2016}. The multicomponent versions of these scalar NLS type equations support bright solitons with interesting shape changing (energy sharing/switching) collisions \cite{RK,Kanna2001,Kanna2010, Vijayajayanthi2008,KANNA2015}. These interesting energy sharing collisions find applications in the context of realizing gates based on soliton collisions \cite{ Jaku1998, Stei2000, Kanna2003,Vijaya_cnsns,  Kanna2018}. However, to date, the intriguing process of soliton interactions remains unexplored in the context of nonparaxial regime except a work that showed a glimpse of the former \cite{WANG2005}. We are hence interested to study the interactions of  bright solitons numerically. The split-step Fourier method based on Feit-Flock algorithm is  adopted here to investigate the interaction between two temporally separated bright solitons in the NNLS equation. To study the scattering dynamics of bright solitons in the NNLS equation, we assume the following two temporal bright solitary pulses with equal amplitudes $\Delta$ ($\approx 1$, in the normalized sense)
\bea\label{seed solution}
\mathbf{\Psi}(0,t)= \mathbf{\Psi}\left(0, t+\Delta t_{0}\right)\exp(i \phi)+ \mathbf{\Psi} \left(0, t-\Delta t_{0}\right),
\eea
where $\mathbf{\Psi}\left(0, t+\Delta t_{0}\right)$ denotes the bright solitary wave solution given by equation~\eqref{21} with   amplitude $\Delta$ $\approx 1$, and $\phi$ indicates an initial phase difference between the two temporally solitary pulses initially separated by a distance  $\Delta t_{0}$. For the simulations performed here, we choose the boundary conditions to minimize the undesired effects such as  reflection of radiation at the boundaries of the computational window.
In what follows, we present the coherent interactions of nonparaxial bright solitons with different parametric choices of obtained solutions and qualitatively discuss the physics behind the interaction dynamics in detail.
\begin{figure}[t]
	\bottominset{(a)~$\bm{\phi=0}$}{\centering\includegraphics[scale=0.32]{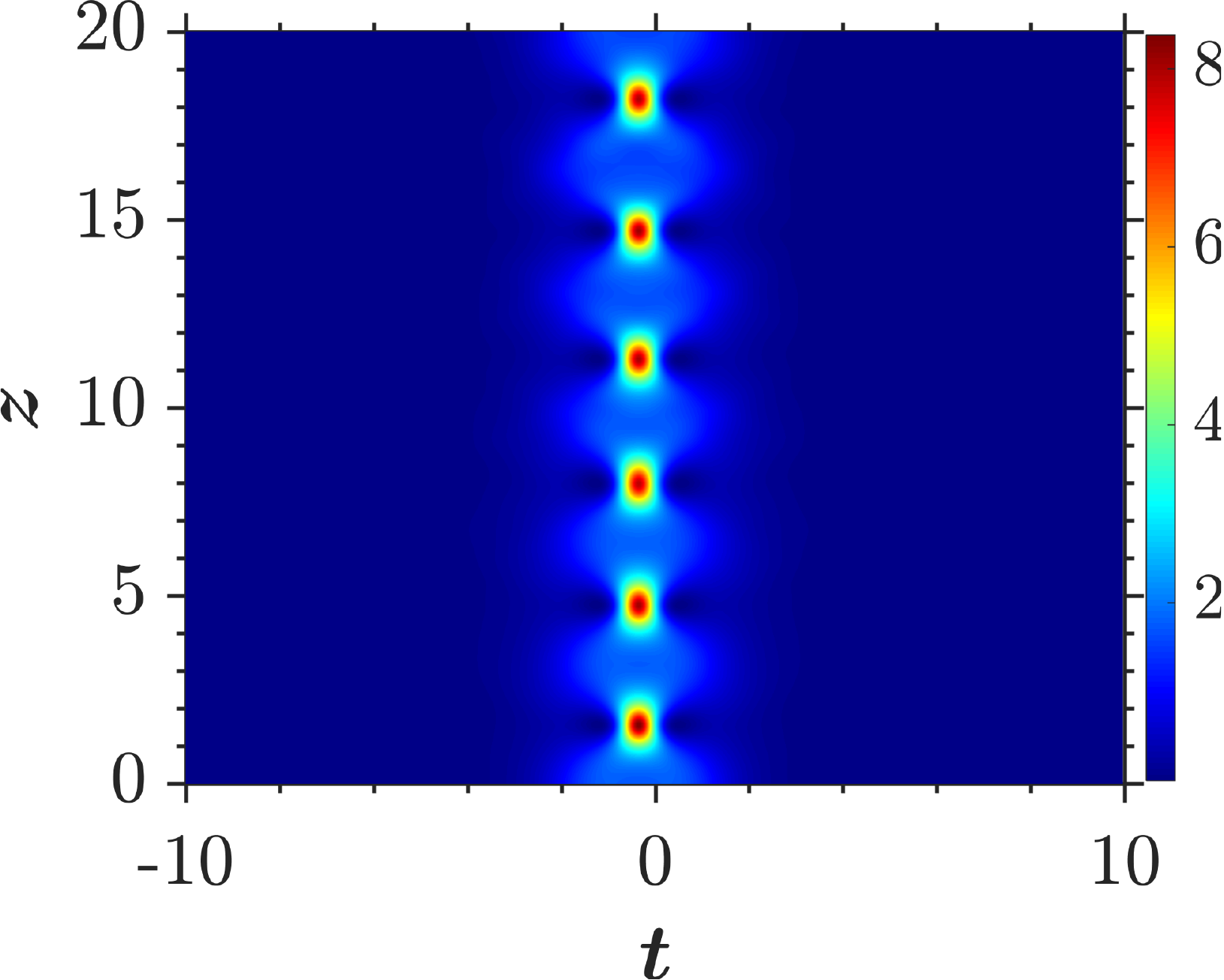}}{-0.2in}{-0.1in}
\bottominset{(b)~$\bm{\phi=\pi/2}$}{\centering\includegraphics[scale=0.32]{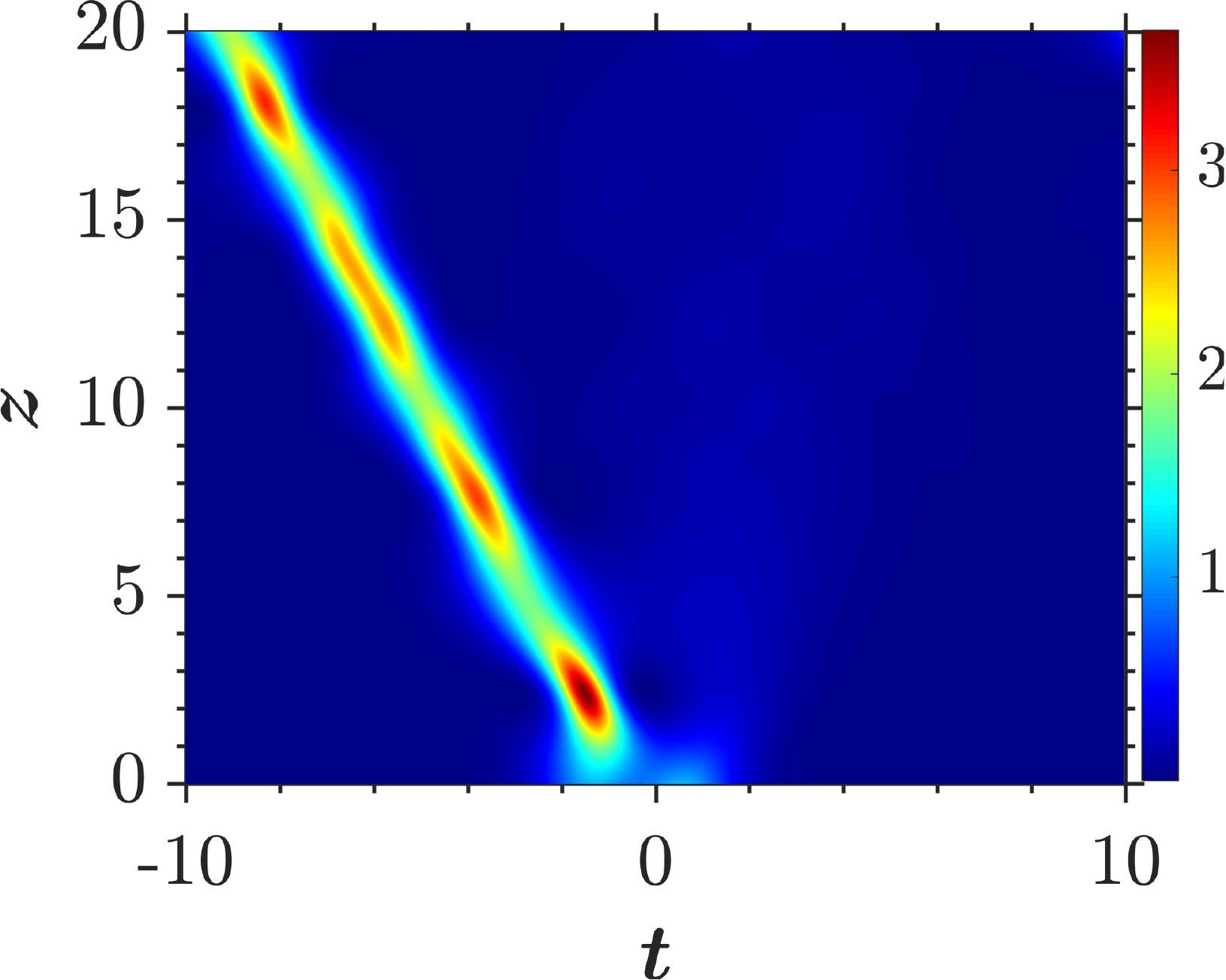}}{-0.2in}{-0.1in}
\bottominset{(c)~$\bm{\phi=\pi}$}{\centering\includegraphics[scale=0.32]{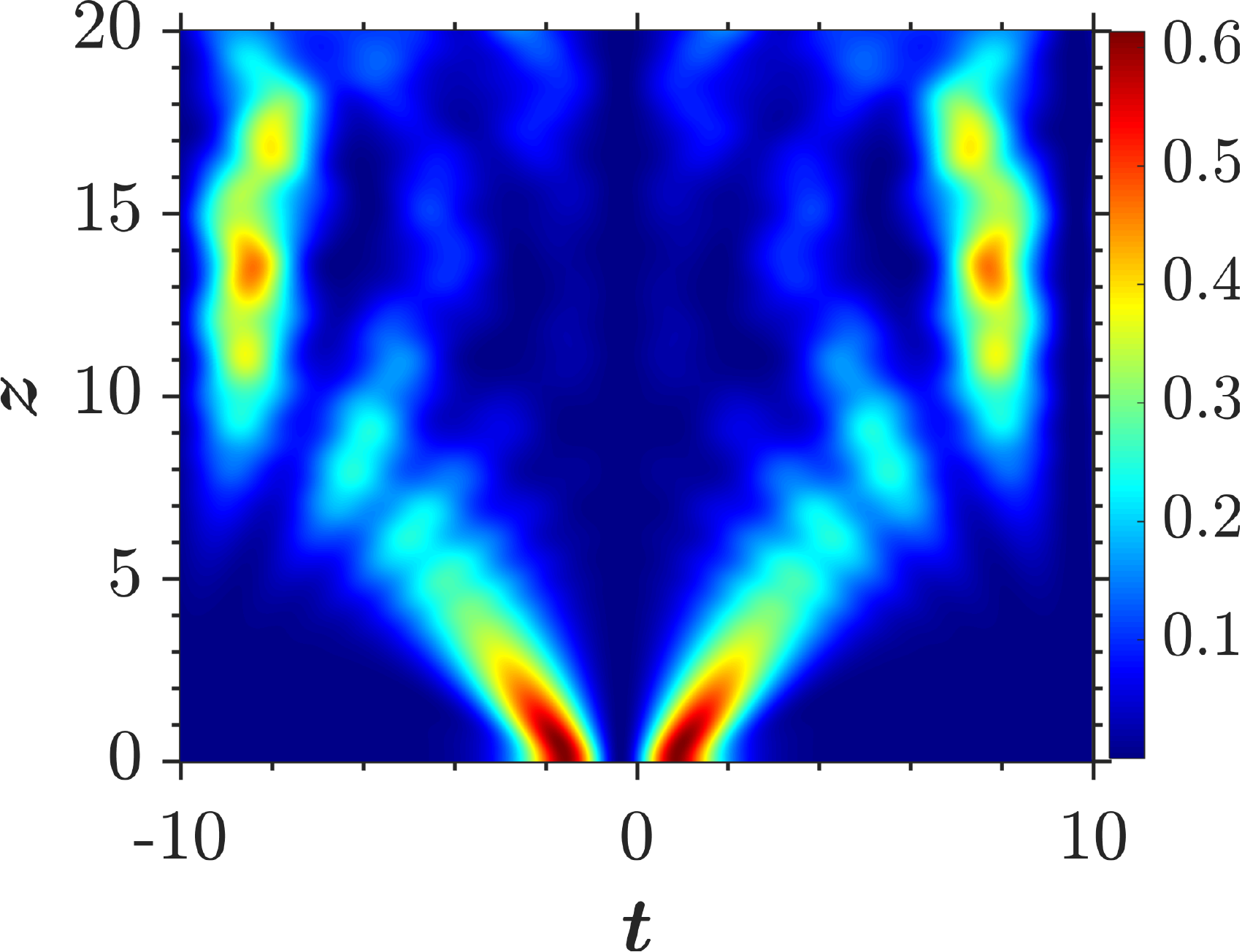}}{-0.2in}{-0.1in}
	\caption{Interaction of dynamics of two bright solitary waves with parameters as $\kappa=0.01$, $\gamma=1$, $\Delta t_{0}=1$, and $b_{1r}=b_{1i}=\alpha=1$. }\label{Numerical1}
\end{figure}

To start with, we consider the collision for the parametric choice $\kappa=0.01$, $\Delta t_{0}=2$, and vary phase from $\phi=0$ to $\pi$ as  presented in Fig.~\ref{Numerical1}. For $\phi=0$, it exhibits an in-phase interaction dynamics and forms oscillating bound solitary waves as shown in Fig.~\ref{Numerical1}(a). Note that, these localized structures maintain their velocity throughout the propagation and retain their shape throughout the medium. The scenario is changed  for the choice of phase $\phi=\pi/2$  as observed in Fig.~\ref{Numerical1}(b), where the bound solitary waves execute oscillations and  deviate away from the central position. Also, the intensities of the interacting pulses are decreased considerably compared to Fig.~\ref{Numerical1}(a), while their widths are extended. The interacting solitary waves experience a significant drift in their path after collision which indicates a strong repulsion between them. This leads to an increase in their separation distance after collision. For the case, $\phi=\pi$,  the interacting pulses become  unstable and dispersion radiations are created [see Fig.~\ref{Numerical1}(c)]. Thus, when the pulses are separated by short distance, stable solitary waves are formed when their phases are correlated \cite{Govind}.

\begin{figure}[t]
	\bottominset{(a)~$\bm{\phi=0}$}{\centering\includegraphics[scale=0.32]{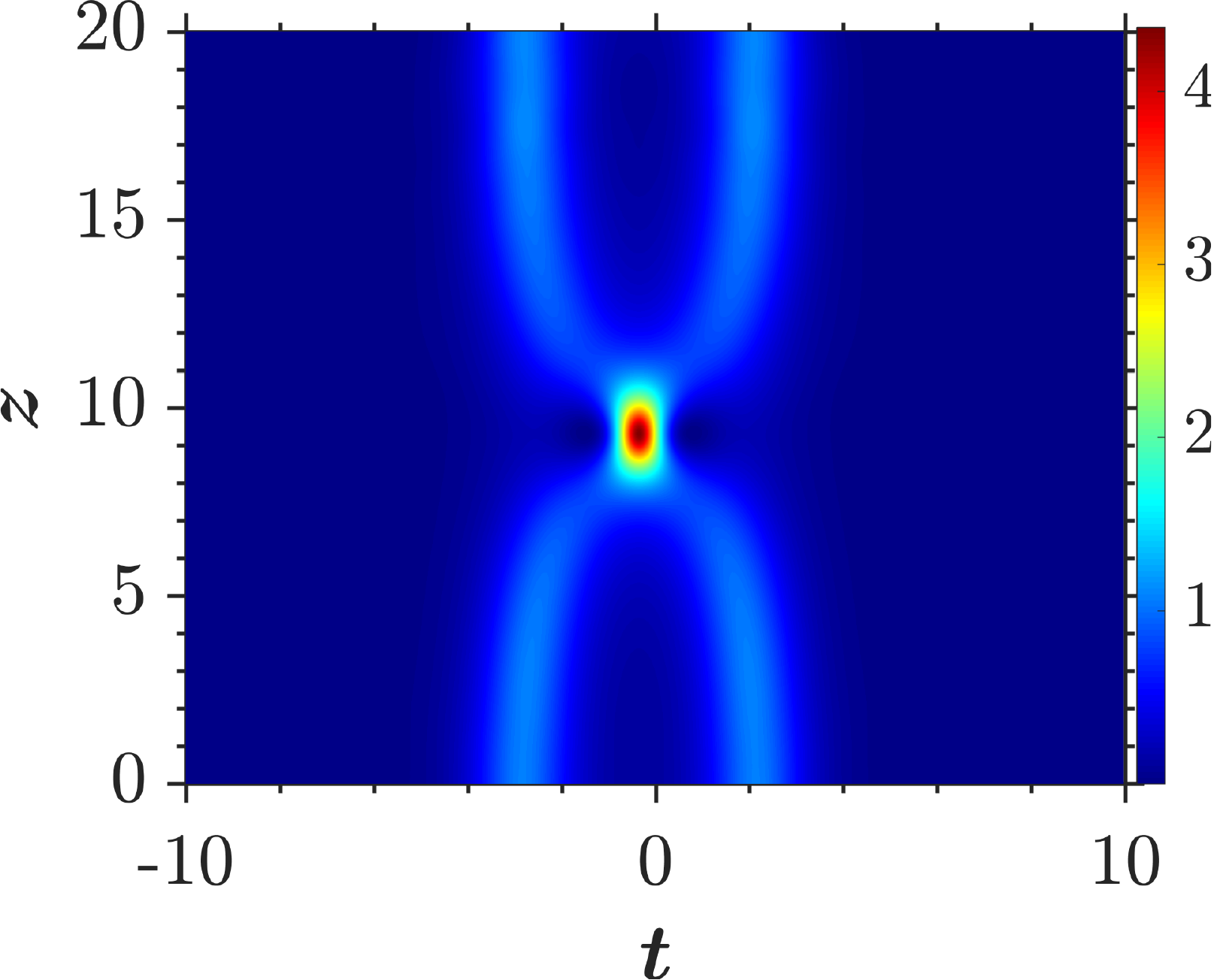}}{-0.2in}{-0.1in}
	\bottominset{(b)~$\bm{\phi=\pi/2$}}{\centering\includegraphics[scale=0.32]{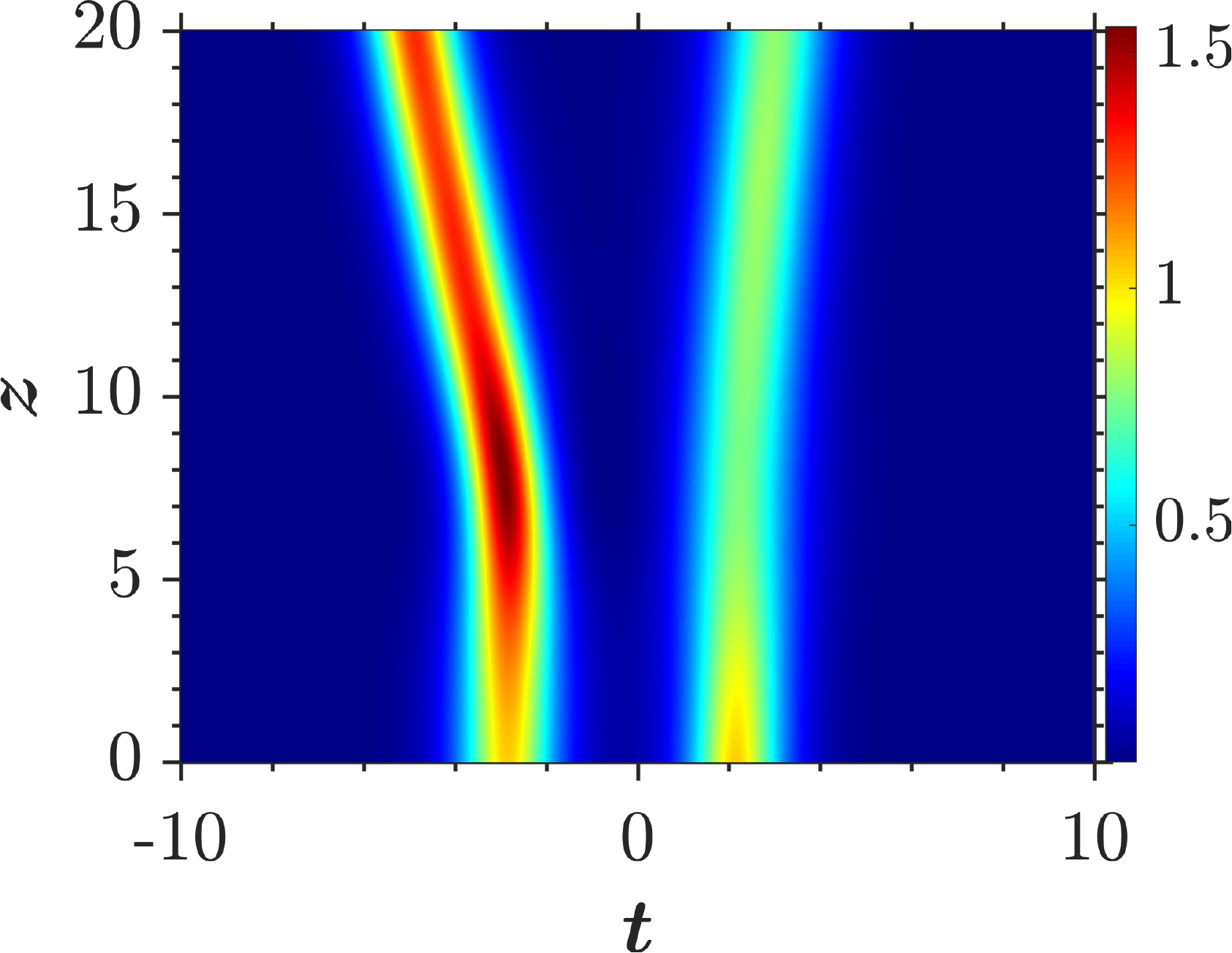}}{-0.2in}{-0.1in}
	\bottominset{(c)~$\bm{\phi=\pi$}}{\centering\includegraphics[scale=0.32]{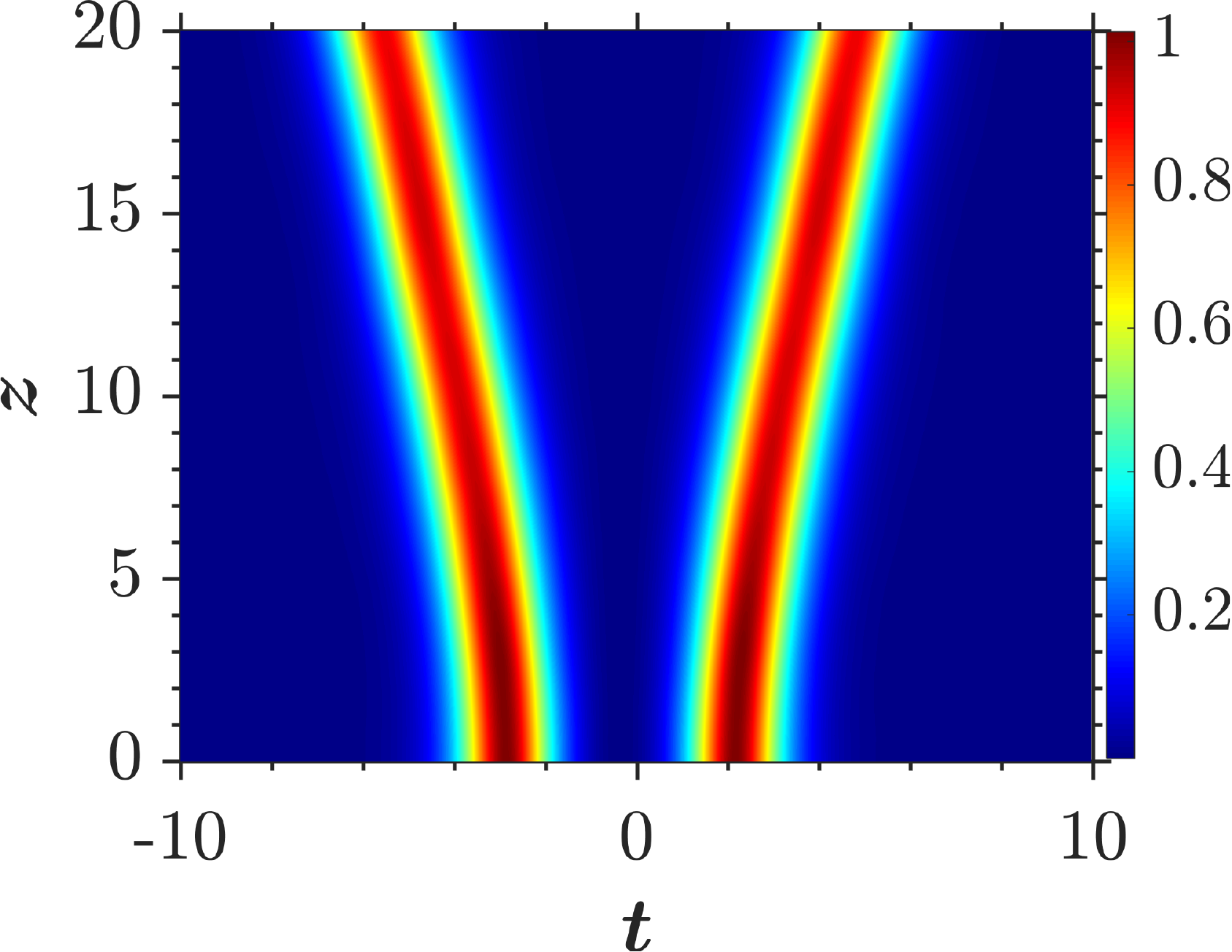}}{-0.2in}{-0.1in}
	\caption{The density plots of interaction of two bright solitary waves. The parameters are the same as in Fig.~\ref{Numerical1} except $\Delta t_{0}=2.5$.}\label{Numerical2}
\end{figure}
\begin{figure}[t]
	\centering
	\bottominset{(a)~$\bm{\phi=0}$}{\centering\includegraphics[scale=0.3]{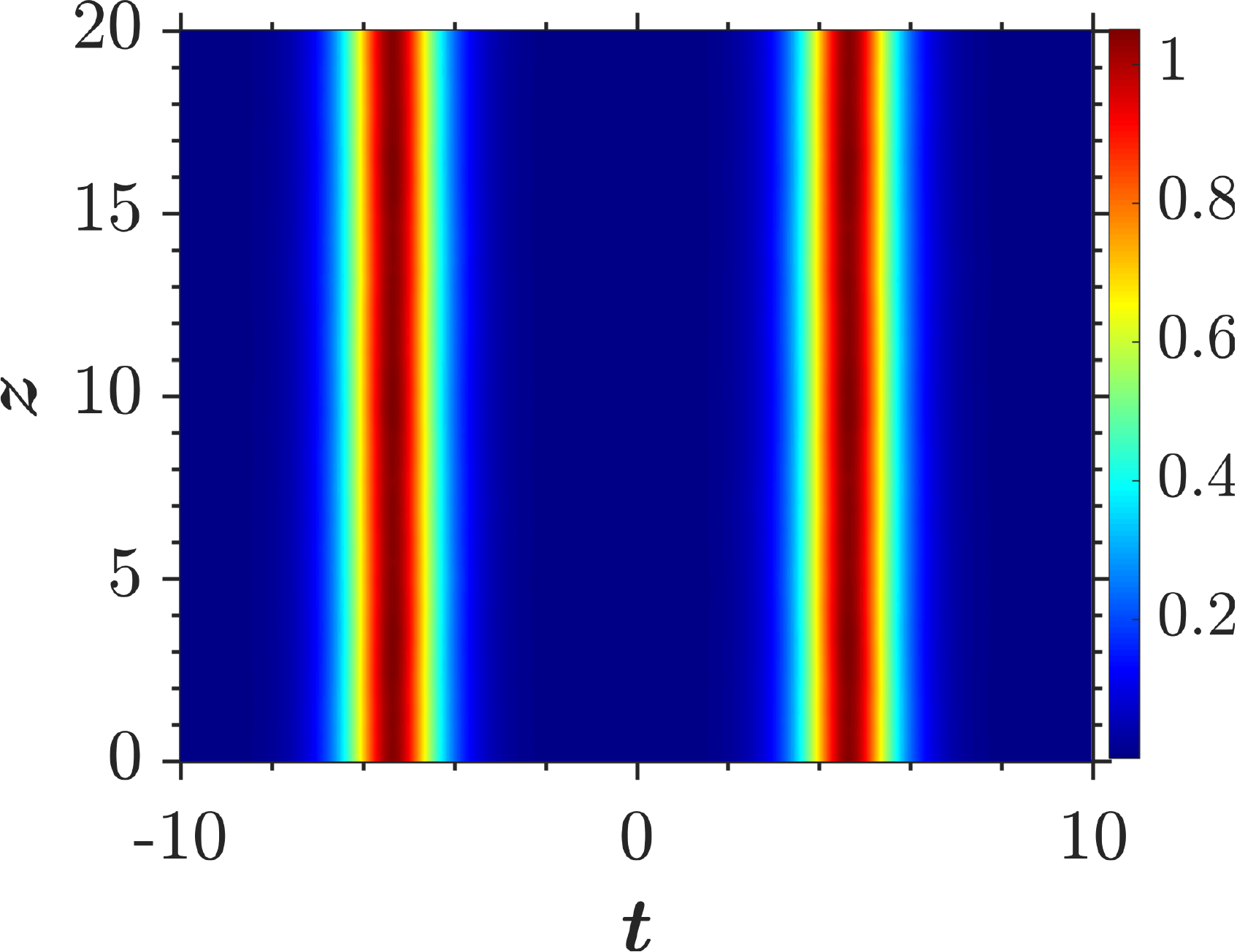}}{-0.2in}{-0.1in}
	\caption{The evolution of interaction of two bright solitary waves. The parameters are the same as in Fig.~\ref{Numerical1} except $\Delta t_{0}=5$.}\label{Numerical3}
\end{figure}

Next, we increase the separation distance between the solitary waves and investigate the collision dynamics for in-phase, out-of-phase, and $\phi=\pi/2$ choices for comparative purpose, as shown in Figs.~\ref{Numerical2}(a)-(c). Surprisingly, in this non-integrable system, we obtain standard soliton like shape-preserving collision for the case $\phi=0$. For $\phi=\pi/2$, there is a non-trivial energy switching among the interacting solitary waves during the collision accompanied by bending (drifting) of the interacting solitary waves resulting an increase in their relative separation distance. For the choice, $\phi=\pi$, also similar behavior takes place [see Figs,~\ref{Numerical2} (a)-(c)]. For larger separation distance there is no passing through collision and we observe only parallel propagation of bound solitons as noticed in Fig.~\ref{Numerical3}(a). It is to be noted that in contrast to the standard NLS and Manakov systems where the solitons exhibit conservation of energy during their collisions, the solitary waves of the present system (1) do not preserve the total energy, rather it conserves the quantity $c_{1}=\int^{+\infty}_{-\infty}\left(|\mathbf{\Psi}|^{2}-i\kappa\left(\mathbf{\Psi^{*}} \frac{\partial \mathbf{\Psi}}{\partial z}-\mathbf{\Psi^{*}} \frac{\partial \mathbf{\Psi^{*}}}{\partial z}\right)\right)dt $. This constant of motion $c_{1}$  can be easily obtained by considering the NNLS equation and its complex conjugate equation and by using the asymptotic behavior of bright solitary waves $\left(\mathbf{\Psi}\xrightarrow[t\rightarrow \pm \infty]{}0\right),$ with a simple algebra. Here the nonparaxial parameter ($\kappa$) is responsible for the violation of the conservation of energy, which is also corroborated through numerical simulations. In particular, as shown in Fig.~5(c), some amount of energy of the solitary wave gets radiated which further proves that the norm of the solitary wave denoting the total intensity is non-conserved.
\begin{figure}[t]
	\bottominset{(a)~$\bm{\phi=0}$}{\centering\includegraphics[scale=0.32]{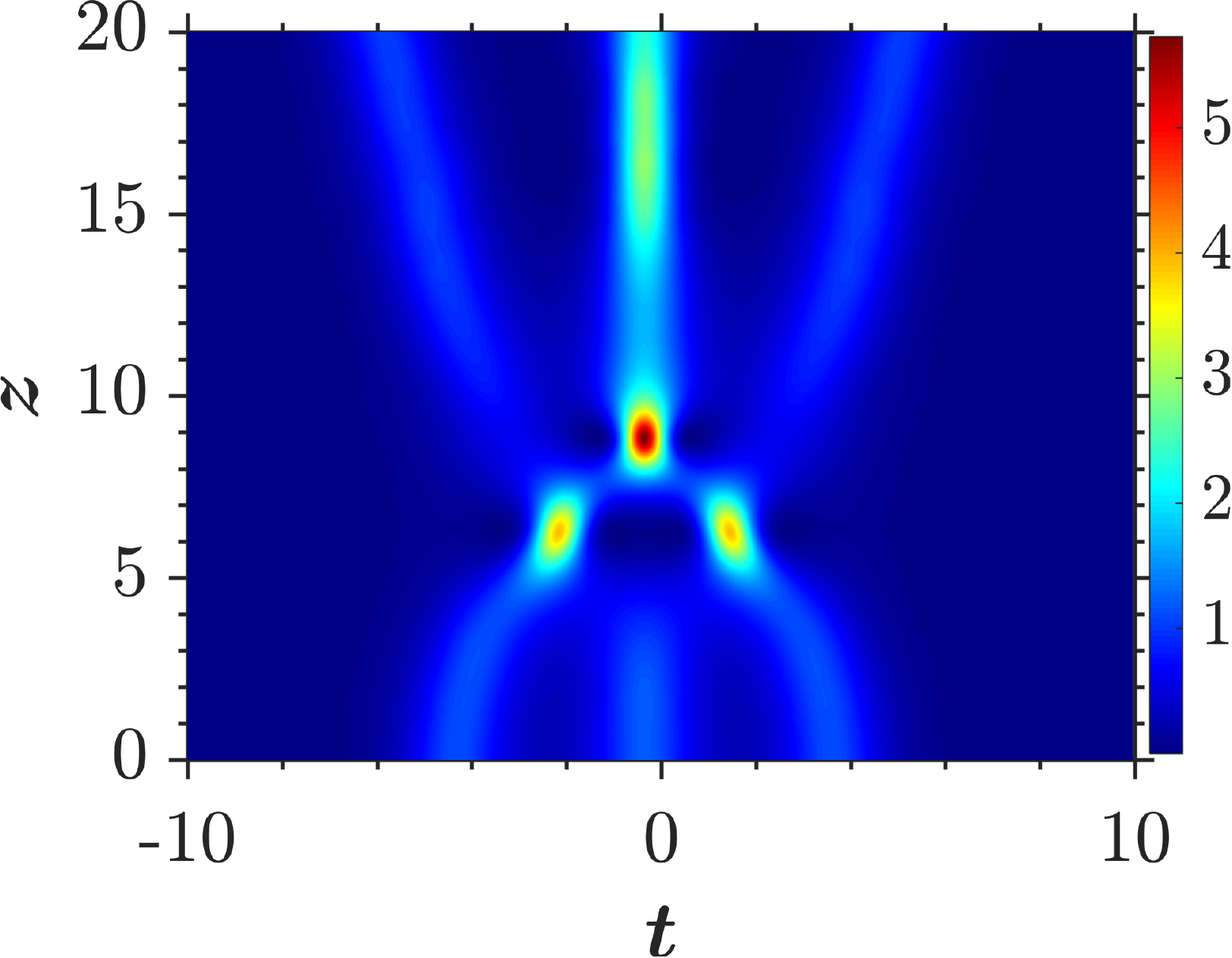}}{-0.2in}{-0.1in}
	\bottominset{(b)~$\bm{\phi=\pi/2$}}{\centering\includegraphics[scale=0.31]{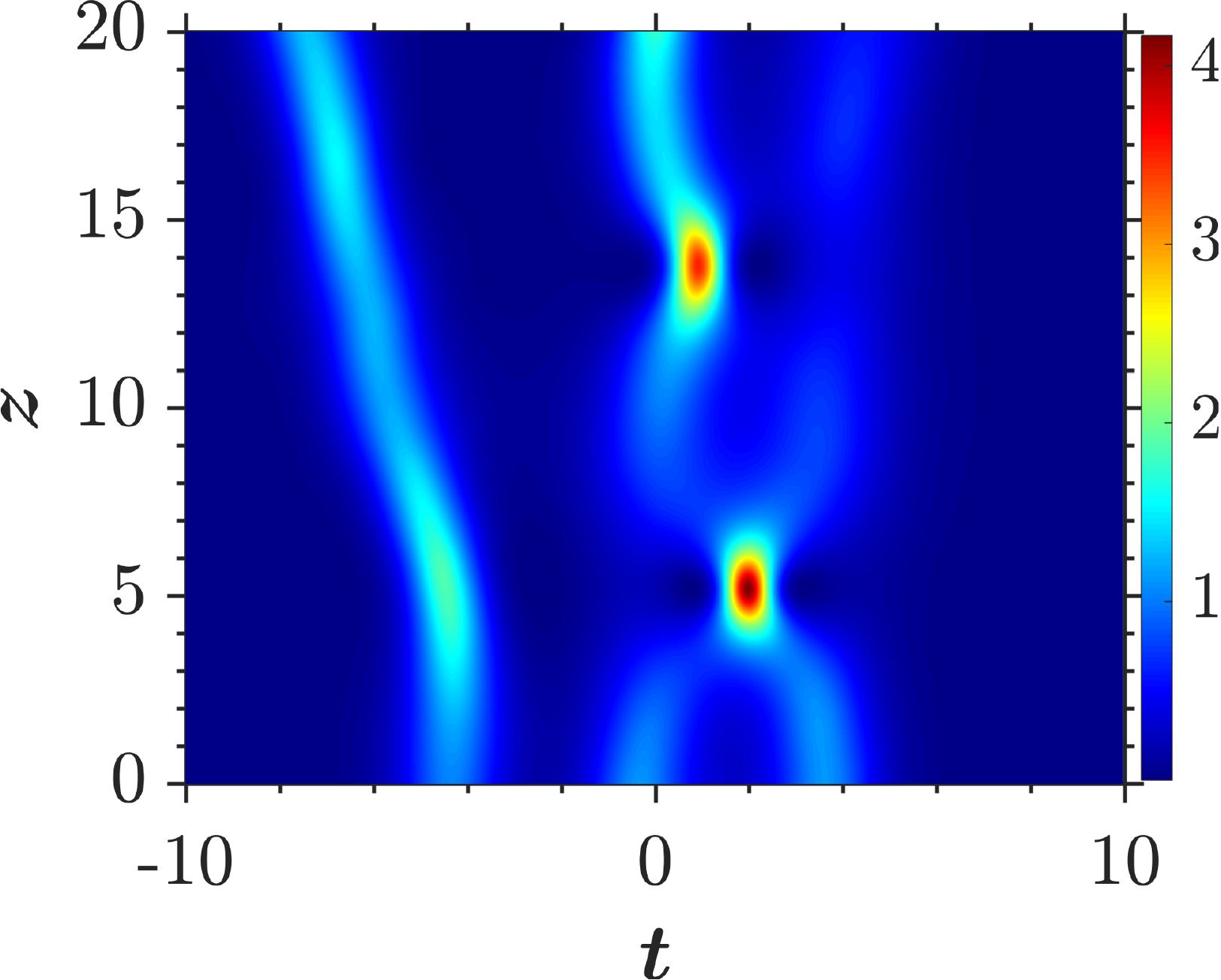}}{-0.2in}{-0.1in}
	\bottominset{(c)~$\bm{\phi=\pi$}}{\centering\includegraphics[scale=0.31]{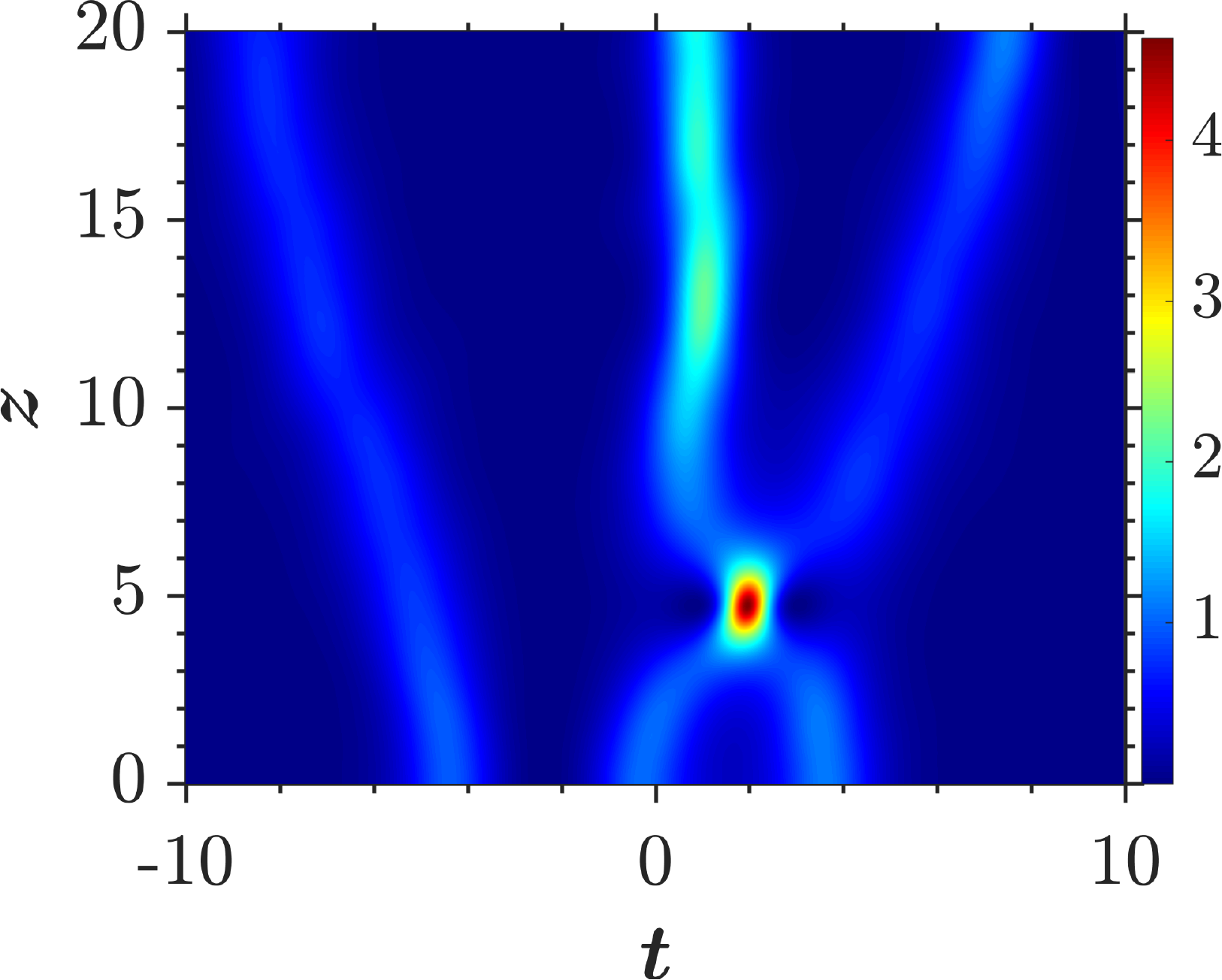}}{-0.2in}{-0.1in}
	\caption{The density plots of interaction of three bright solitary pulses. The parameters are the same as in Fig.~\ref{Numerical1} except $\Delta t_{0}=4$.}\label{Numerical4}
\end{figure}

Finally, it is interesting to reveal the scattering nature of three bright solitary waves by altering the phase of solitary waves from in-phase (zero) to out-of-phase ($\pi$) as shown in Fig.~\ref{Numerical4}. In the case of in-phase, three solitary waves are attracted to each other at $z\sim7$ ($z$ denotes the propagation distance of the medium), afterwards they  get separated symmetrically to each other and also retain their shape after collision (see Fig.~\ref{Numerical4}(a)). We also observe a non-trivial energy switching in the second (middle) solitary wave. By tuning the phase to $\pi/2$, we identify an  interesting interaction dynamics, where one solitary pulse (left side) is completely separated from other solitary waves and is deviated away from the remaining solitary waves. The rest of the two solitary pulses initially propagate within a very short separation distance and after the collision due to repulsion between the solitary pulses the separation distance between them is increased. In particular, two solitary pulses have  distinct intensity profiles, featuring an energy transfer from one solitary pulse to another one as presented in Fig.~\ref{Numerical4}(b). A similar
collision behavior with a significant energy switching in the right most solitary wave (before collision) can be observed for the case $\phi=\pi$ with a slight modification as given in Fig.~\ref{Numerical4}(c). Note that after collision,  there is a corresponding decrease in the intensities of other two solitary pulses.
\section{Conclusion}
To conclude, we have investigated the integrability aspects of the NNLS equation by employing the Painlev\'e singularity structure analysis. Based on this analysis, we have proven that the NNLS equation fails to satisfy the Painlev\'e test as it is not free from the movable singularity at the resonance $j=3$. Nevertheless, we have then constructed bright solitary wave for the NNLS equation by using the Hirota's bilinearization method. We have demonstrated stable propagation over long distance even in the presence of external perturbation, which is seeded in the form of a white noise, by employing the SSFM. The scattering dynamics of bright solitary waves has also been investigated by numerical simulation  for different values of separation distance and relative phase. This numerical study reveals that there is an energy/intensity switching among the colliding solitons  after collision, due to the nonparaxiality/spatial dispersion. Also, the collision leads to a stronger repulsion between the solitons which results in an increase in the separation distance between the solitons after interaction. Ultimately, there is a significant deviation in the trajectory of solitary wave. We anticipate that the results will shed  light in the formation, propagation and collision of solitary waves in nonparaxial nonlinear  media. The energy switching phenomenon during collision in the NNLS  system can find applications in optical switching devices, beam steering and in soliton collision based optical computing.

\section*{Acknowledgement}

The work of K T is supported by a Senior Research Fellowship from Rajiv Gandhi National Fellowship (Grant No. F1-17.1/2016-17/R GNF-2015-17-SC -TAM-8989), University Grants Commission (UGC), Government of India. The work of T K was supported by the Science and Engineering Research Board, Department of
Science and Technology (DST-SERB), Government of India, in the form of a major research
project (File No. EMR/2015/ 001408). A G acknowledges the support of DST-SERB for providing a Distinguished Fellowship (Grant No. SB/DF/04/2017) to M L in which A G was a Visiting Scientist. A G is now supported by University Grants Commission (UGC), Government of India, through a Dr. D. S. Kothari Postdoctoral Fellowship (Grant No. F.4-2/2006 (BSR)/PH/19-20/0025).

\section*{References}

\bibliography{reference}

\end{document}